\documentclass[%
groupedaddress,
 amsmath,amssymb,
 aps,
prl,
twocolumn,
floatfix,
notitlepage
]{revtex4-1}

\usepackage{graphicx}% Include figure files
\usepackage{dcolumn}% Align table columns on decimal point
\usepackage{bm}% bold math
\usepackage{subfigure}%subfigure enviroment
\makeatletter
\renewcommand{\thesubfigure}{\alph{subfigure}}% (a) -> a
\renewcommand{\@thesubfigure}{\thesubfigure)\hskip\subfiglabelskip}% a -> a)
\makeatother
\usepackage[colorlinks=true,bookmarks=true,%
     pdfborder={0 0 0},linkcolor=blue,urlcolor=blue,%
     breaklinks=true,bookmarksnumbered=true]{hyperref}
\usepackage{subfigure}%subfigure enviroment
\makeatletter
\renewcommand{\thesubfigure}{\alph{subfigure}} %(a) -> a
\renewcommand{\@thesubfigure}{\thesubfigure)\hskip\subfiglabelskip} %a -> a)
\makeatother
\usepackage[stable]{footmisc}
\usepackage{wasysym}
\usepackage{times}

\def \x{\bm{x}}

\def \S{\bm{S}}

\def \k{\bm{k}}

\def \pt{\partial}

\def \L{\mathcal{L}}

\def \si{\sigma}

\newcommand{\abs}[1]{\lvert#1\rvert}

\newcommand{\meanvalue}[3]{\mbox{$\langle #1 | #2 | #3 \rangle$}}

\newcommand{\ev}[1]{\mbox{$\langle #1 \rangle$}}

\newcommand{\ket}[1]{\mbox{$| #1 \rangle$}}

\newcommand{\h}[1]{\hat{#1}}

\newcommand{\dwx}[1]{\textcolor{blue}{#1}}

\newcommand{\eqdisp}[1]{Eq.~(\ref{#1})}

\def \dwx{\color{black}}
\begin{document}

\title{Local Density of States {induced near} Impurities in Mott Insulators
}

% Force line breaks with \\

\author{Wenxin Ding$^{1,2,3}$}\email{wenxinding@gmail.com}
\author{Qimiao Si$^2$}
\affiliation{%
  $^1$School of Physics and Optoelectronics Engineering, Anhui University, Hefei, Anhui Province, 230601, China\\
  $^2$Department of Physics \& Astronomy, Rice University, Houston, Texas 77005, USA\\
  $^3$Kavli Institute for Theoretical Sciences,\\
   University of Chinese Academy of Sciences, Beijing, 100049, China
}

\date{\today}% It is always \today, today,
             %  but any date may be explicitly specified{\dwx we study the impurity}

\begin{abstract}
%Recent scanning tunneling microscope measurements \qs{have} probed
The local density of states near dopants or impurities has recently been probed by scanning tunneling microscopy in both the parent and very lightly doped compounds of the high-$T_c$ cuprate superconductors. {  We use a slave-rotor representation of the Hubbard model to compute the local density of states on impurities in a Mott insulator. Our calculation accounts for} the following key features of the experimental observation: i) positions and amplitudes of the in-gap spectral weights of a single impurity; ii) the spectral weight transfer from the upper Hubbard band to the lower Hubbard band; iii) the difference between the cases of single and multiple impurities. For multiple impurities, our study explains the complete suppression of spectral weight observed at precisely the Fermi energy and links this property to zeros of the underlying bulk Green's function of the Mott insulating phase.
\end{abstract}

\pacs{Valid PACS appear here}% PACS, the Physics and Astronomy
\maketitle
%\tableofcontents

%\section{Introduction}
{\it Introduction.}~~
The high-$T_c$ cuprate superconductors are generally interpreted as doped Mott insulators \cite{Lee2006}.
 The undoped parent compounds are ordered antiferromagnetically.
 Although the parent compounds are
 %genuinely
 three-band, charge transfer insulators \cite{Zaanen1985}, it is believed that the single-band Hubbard model (HM) provides an adequate effective description.
% of the system.
%In the undoped compounds, t
 The strong, onsite Coulomb repulsion, Hubbard-$U$,
 forbids two electrons to occupy the same Cu site,
% thus creating
 {thereby creating}
 a Mott gap $\propto U$. {  As a result, the electronic spectrum splits} into %the
 an upper Hubbard band (UHB) and a lower Hubbard band (LHB).
%The localized electrons form local moments which interact antiferromagnetically \cite{Anderson1950}, leading to the antiferromagnetic (AFM) ground state. Subsequent s
Superconductivity arises from
% further
 carrier-doping  the parent compounds beyond a threshold concentration.

The rich properties of the doped cuprates suggest the importance of studying Mott insulators {  with} dilute dopants or impurities. The local density of states (LDOS) near dopants or impurities in the parent Mott insulator has been studied through scanning tunneling microscope (STM) measurements \cite{Ye2012b,Cai2015}. These experiments have uncovered a number of features on the electronic excitation spectrum of a Mott insulator.
 For a single impurity, the in-gap states emerge from the UHB above the Fermi energy. When impurity concentration increases, the in-gap states gradually fill up the Mott gap, but a ``V''-shaped dip forms near the Fermi energy.
 %This dip indicates
  {The observed dip means} that the impurities or dopants
 cannot produce in-gap states exactly at the Fermi energy.
 %Furthermore, the in-gap states closer to the Fermi energy always have smaller spectral weights.

The experimental development is exciting as it bridges between the clean parent compounds and the heavily debated, lightly doped but metallic pseudogap regimes \cite{Keimer2015}.
A systematic study of the Mott insulator in the presence of a single defect dopant or impurity is highly desirable. Previous efforts on this type of problems \cite{Liu1992,Poilblanc1994,Leong2014} have been mostly numerical, which are restricted by finite-size effect, and have yet to achieve the understanding of the key experimental observations mentioned earlier.

In this Letter, we study the LDOS of single and multiple impurities in a Mott insulator based on a slave rotor representation of the HM in the thermodynamic limit. We find clear, impurity-induced in-gap bound states, descending from the UHB as observed in Ref. \cite{Ye2012b} {  in Ca$_2$CuO$_2$Cl$_2$(CCOC)}. In addition, we obtain the correct spectral weight transfer from the UHB to the LHB. Systematic calculations of the bound state energies and their corresponding spectral weights provide qualitative understanding about the experiments of Ref.~\cite{Cai2015} {  which is done for Bi$_2$Sr$_{2-x}$La$_x$CuO$_{ 6+\delta}$(La-Bi2201) at two hole densities $p=0.03$ and $0.07$}: i) the bound states cannot reach the Fermi energy; ii) the bound states with energies closer to the Fermi energy have smaller spectral weights. We further show that the vanishing of the LDOS at the Fermi energy reflects the zeros of the Green's function, i.e the Luttinger surface \cite{Dzyaloshinskii2003}, of the underlying Mott insulator. That is a feature of considerable interest to the Mott insulator per se and to the physics of the pseudogap regime of the underdoped cuprates  \cite{Konik2006, KYYang2006, Dave2013}. {  Recently, similar LDOS on apical oxygen impurities are also observed in a different Mott insulator Sr$_2$IrO$_4$ \cite{Sun2019}.}

%The rest of the letter is organized as the follows. First, we briefly review the slave rotor formalism, and its mean field solution of the bulk Mott insulator. Then we discuss the effective rotor action in the presence of impurity potentials. Subsequently, both the rotor spectral functions and electronic LDOS of a single impurity are computed numerically. Thereafter, we discuss the multi-impurity solution, and its implication for the finite doping measurements. Finally, we close by discussions of implications of our work.

%\section{The Slave Rotor Formalism}

{\it The slave-rotor approach.~~}
We consider the single band Hubbard model on a square lattice
\begin{equation}
  \label{eq:Hubbard-model}
  H_{HM} = \sum_{i} H_{at}(i) - \sum_{ij,\si} (t_{ij} d^{\dagger}_{i\si} d_{j\si} + H.c.),
\end{equation}
in which $H_{at}(i) = \frac{U}{2}(n_{\uparrow} - 1/2 ) (n_{\downarrow} - 1/2 )$.
For simplicity, we consider only hopping between nearest-neighbor ($nn$) sites, $\ev{ij}$.
The full energy spectrum of $H_{at}(i)$ can be economically represented by a rotor kinetic energy $H_{at}(i) \rightarrow U \hat{L}_{i}^2 /2$ \cite{Florens2002,Florens2004} with $\hat{L}_{i} = -i\pt_{\theta_i}$, which provides a tractable reference point for perturbative expansion in $t/U$. In this slave-rotor representation, the bare electron operator is written as a product of the rotor field and a fermionic spinon operator $d_{i\alpha} \equiv f_{i\alpha} e^{-i \theta_i}$,
with the constraint
\begin{align}\label{eq:constrain-h}
 \hat{L_i} = \sum_{\si}(f^{\dagger}_{i\si} f_{i\si} -1/2).
\end{align}
In place of the phase field  one could work with the complex field
$e^{i\theta_i} = X_i$, with the additional constraint $\abs{X_i}^2 = 1$.
The two constraints are enforced by introducing two Lagrangian multipliers, $h_i$ and $\lambda_i$.
{  In terms of the fermionic $f_i$ and  complex rotor $X_i$ operators, the physical $d_i$-electron operator at site $i$ is
expressed as follows:
\begin{equation}
  \label{eq:d_by_fX}
  d_{i\si} \equiv f_{i\si} X_i^{*} \, .
\end{equation}
Correspondingly, the rotor operator becomes $\hat{L}_i = (h - X_i^* \pt_t X_i)/U.$

Intuitively, in the Mott phase, the rotor fields describe the bosonic Hubbard-$U$ charge dynamics while the $f$-spinons describe the magnetic fluctuations. It is shown in Ref.~\cite{Ding2014} that by integrating out the $X$-fields, the Heisenberg interaction term is correctly recovered with the physical spins given by the $f$-spinons as $\S_i = \sum_{\alpha,\beta} f^\dagger_\alpha \bm{\si}_{\alpha \beta} f_\beta$.}

A saddle point solution \cite{Florens2004,Lee2005} (see Supplemental Material \cite{SM}, Eq.~(\ref{eq:mf-L}) for details) is found {for a paramagnetic Mott insulating phase.
%Due to the spin-charge separation nature of the Mott phase,
This saddle point state is described by a free $f$-spinon theory and a free $X$-field theory. Their corresponding Green's functions are
\begin{align}
  G_{f,0}(\omega;\k) = (\omega + h -Q_f \epsilon_{\k})^{-1}, \\
 G_{X,0}(\nu;\k) = (- \nu^2/U + 2ih\nu/U + \lambda + Q_X \epsilon_{\k})^{-1},
\end{align}
where $\epsilon_{\k} = -2 t (\cos k_x + \cos k_y)$ is the bare lattice dispersion function, $Q_{f}  = Q_{f\ev{ij}}= \ev{ X_j^* X_i},~Q_{X \ev{ij}}  = \ev{\sum_{\si} f_{j \si}^\dagger f_{i \si}}$.
The spin and charge sectors are coupled through the self-consistency between $Q_f$ and $Q_X$.
{\dwx Although we only consider the $nn$ hopping in this work, the inclusion of other hopping terms such as the next-nearest-neighbor term would only renormalize $Q_{f}$ and $Q_x$. %within the current computation scheme.
}
According to Eq. (\ref{eq:d_by_fX})}, the electronic Green's function is calculated via the rotor and spinon Green's functions according to
\begin{equation}
 i G_d(t;\x,\x') =  - G_f(t;\x,\x')  G_X(-t;\x,\x'). \label{eq:Gd}
\end{equation}
%We are interested in
We focus on the electronic LDOS $\rho_d(\omega;\x) = - (\pi)^{-1} \Im m[G^R_d(\omega;\x,\x)]$, expressed as
\begin{equation}\label{eq:Gd-conv}
  \begin{split}
   & \rho_d(\omega;\x)
   = \int d\omega' \rho_f(\omega';\x) \rho_X(\omega - \omega';\x)\\
   & \times (n_f(\omega') + n_B(\omega - \omega')),
 \end{split}
  \end{equation}
where $ \rho_f[\omega;\x] = - (\pi)^{-1 } \Im m[G^R_f(\omega)]$, and $\rho_X[\omega ] =- (\pi)^{-1 } \Im m[G^R_X(\omega)]$.

%The bulk Mott insulating phase is described at the saddle point level.

% $Q_X = \ev{\sum_\sigma f_{i \sigma}^\dagger f_{j\sigma}}$ is a constant independent of the effective spinon hopping $Q_f$, and, thus self-consistency is trivially satisfied.
In this work, we are interested in the large-$U$ limit {  in which $\lambda \simeq U/4~$ \cite{Ding2014}}.
At the saddle point, $\Im m[G_f(\omega')]$ centers at $\omega' = 0$ with a bandwidth $W_f = 4D~t~Q_f$, which is small because $t~Q_f \ll U$; any impurity effects on $\Im m[G_f(\omega')]$ {\it per se} would be on the same scale.
In the convolution, $\Im m[G_f(\omega')]$ can be regarded as a broadened $\delta$-function since $Q_f \ll U$.
The Mott gap is primarily determined by $\Im m[G_X(\nu)]$. Therefore, the impurity-induced features of the electronic LDOS are mainly determined by the rotor fields,
which shall be our focus in the following.
%section.

{\it The induced rotor impurity potential.}~~

We consider the case of a single impurity in {  CCOC} as experimentally studied in Ref.~\cite{Ye2012b}, which is either a missing chlorine(Cl$^{-}$) {ion} or a calcium(Ca$^{2+}$) {defect}. The vacancy is charged and creates an impurity potential. We model it by a localized, onsite potential $V(\x_i)$, where $\x_i$ denotes the vacancy position.

Consider a single  on-site impurity at $\x_0$:
\begin{equation}
H_1 = V \sum_\sigma d^\dagger_{\x_0,\sigma} d_{\x_0,\sigma} = V \sum_\sigma f^\dagger_{\x_0,\sigma} f_{\x_0,\sigma}.\label{eq:H1}
\end{equation}
{  This bare impurity potential only couples to the spinons and induces a variation of the local spinon density $\delta n_f(\x_0) = \delta\ev{\sum_\sigma f^\dagger_{\x_0\sigma} f_{\x_0 \sigma}}$. However, }
the rotors and the spinons are subject to the constraint Eq. (\ref{eq:constrain-h}).
Through the constraint, the rotors will be perturbed by the impurity potential as well.

In the large-$U$ case, it suffices to solve the constraint in the atomic limit. For the bulk state, we have $\ev{\hat{L}} = 0 \rightarrow h = 0$. For arbitrary $h$, we have
\begin{equation}\label{eq:evL}
  \begin{split}
   &  \ev{\hat{L}} = h/U - \frac{1}{2} \frac{h}{\sqrt{- h^2 + \lambda U}}.
  \end{split}
\end{equation}
% Note that we do not take derivative on the $\lambda X^2$ term. Its contribution should vanish as we can integrate out $\lambda$ and replace it by a $\delta$-function. Then after taking the functional derivative and integration by part, we obtain a term that is odd in frequency.
Therefore, {  $\delta n_f(\x_0)$ further} induces {  a} variation of the Lagrangian multiplier $h$ through the constraint which we shall label as $\delta h(\x_0)$. Using the atomic limit result, $\delta h(\x_0)$ is obtained by solving the following equation:
\begin{equation}
 \frac{\delta h(\x_0)}{U} - \frac{\delta h(\x_0)}{2\sqrt{-\delta h(\x_0)^2 + \lambda U}} =  \delta \ev{n_f(\x_0)}.
\end{equation}

\begin{figure}
  \centering
  % \subfigure[]{\label{fig:rotor-bulk}\includegraphics[width=0.85\columnwidth]{./G0-spectrum.pdf}}
  \subfigure[]{\label{fig:dh-dnf}\includegraphics[width=0.45\columnwidth]{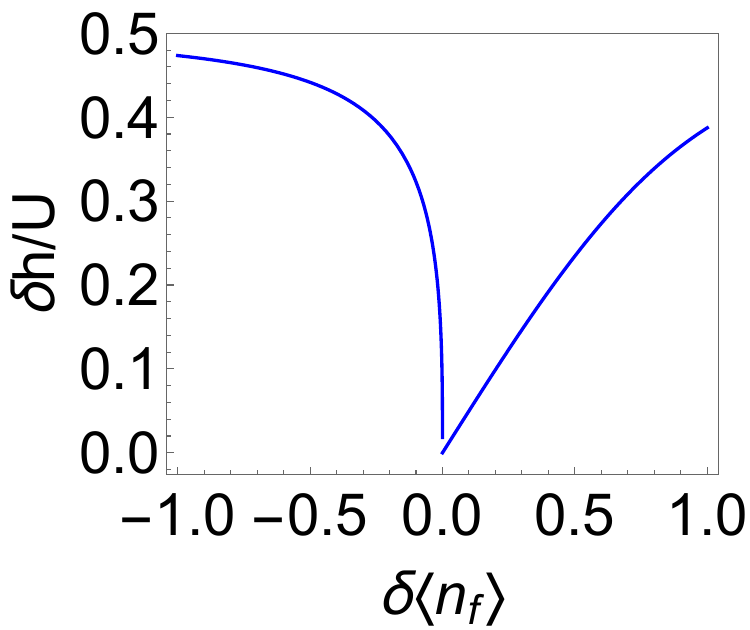}}
  \subfigure[]{\label{fig:rotor-impurity}\includegraphics[width=0.5\columnwidth]{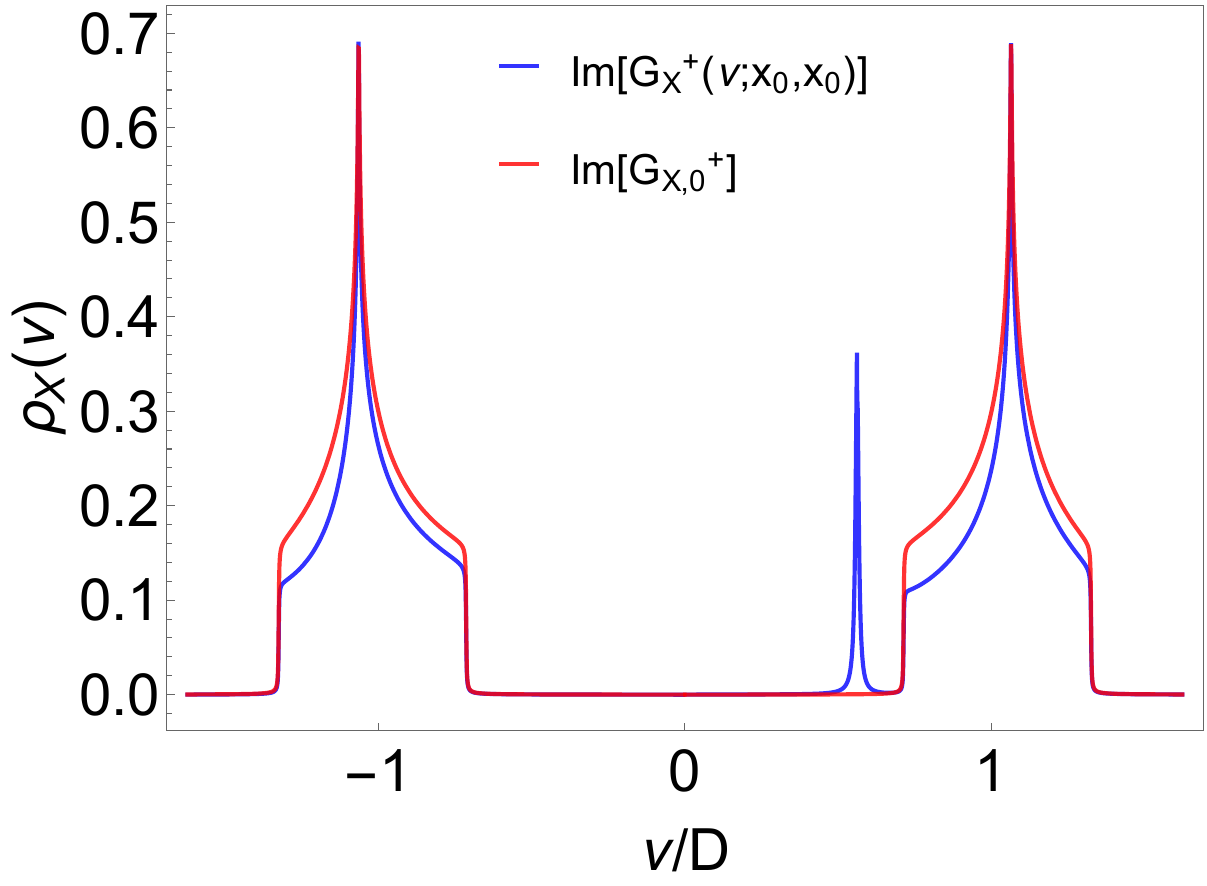}}
    \caption{(Color online) (\ref{fig:dh-dnf}) the solution of $\delta h(x_0)$ plotted as a function of $\delta \ev{n_f(x_0)}$; (\ref{fig:rotor-impurity}) the bulk rotor spectral function {  $\Im m[G_{X,0}^+(\omega;\x,\x)]$ (blue)} compared with that on a single impurity, $\Im m[G_{X}^+(\omega;\x_0,\x_0)]$ {  (red)} .%the real part $\Re e[G_{X,0}^+(\omega;\x,\x)]$(blue) and imaginary part $\Im m[G_{X,0}^+(\omega;\x,\x)]$(orange), and the comparison of $\Im m[G_{X,0}^+(\omega;\x,\x)]$(orange) and that on a single impurity, $\Im m[G_{X,0}^+(\omega;\x_0,\x_0)]$(blue).}\label{fig:rotor-spectrum}
    }%\label{fig:rotor-impurity}
\end{figure}

Taking $\lambda = U/4$, we plot the solution of $\delta h(\x_0)$ as a function of $\delta \ev{n_f(\x_0)}$ in Fig.~(\ref{fig:dh-dnf}).
To solve for the impurity states, we
%assume
 {take}
 $\delta \ev{n_f(\x_0)} =  - 1$, since
 % in experiments
  $V$  {of relevance to the experiments}
   is on the order of $eV$, i.e.
   %sufficiently large
    {much larger than the}
   % comparing to
    spinons' bandwidth. This solution gives us the upper bound of $h(\x_0)$. For the rest of this work, we shall use $\delta h(\x_0) \simeq 0.473 U$ unless specified otherwise.
% (the spinon bandwidth is usually much smaller than the bare electronic energy scale, i.e. $Q_X \ll t$)
%Therefore, we obtain
%\begin{equation}\label{eq:h0}
%  \delta h(\x_0) \simeq
%  \begin{cases}
%    0.473 U, \qquad \delta \ev{n_f(\x_0)} = - 1,\\
%    0.387 U, \qquad \delta \ev{n_f(\x_0)} = 1.
%  \end{cases}
%\end{equation}
Even though negative solutions of $\delta h(\x_0)$ are allowed here, they do not induce in-gap bound states, and thus shall be ignored.
Then the {  rotor} impurity potential due to $\delta h(\x_0)$ is
\begin{equation}
  \begin{split}
    & \h{H}_{X,\x_0} = i h_0(X^*_{\x_0}\pt_t X_{\x_0} - X_{\x_0}\pt_t X_{\x_0}^*)
%\ket{\x_0 }-i\pt_t\bra{\x_0 } \\ & = \int    \frac{d\nu d^2\k d^2\k'}{(2 \pi)^5} i \nu \delta h(\x_0) \ket{\nu    \k}\bra{\nu \k'},
  \end{split}\label{eq:imp-pot}
\end{equation}
where we label $h_0 = 2 \delta h(\x_0)/U$.

%\subsection{Rotor Spectral Function}
{\it %The T-matrix calculation of the impurity
 {Impurity} state Green's function.}~~
%In this section, we use the $T$-matrix formalism to perturbatively calculate the LDOS of a single impurity in a Mott insulator.
 {We now turn to calculate the LDOS of a single impurity in a Mott insulator
using the $T$-matrix formalism.}
%In principle, as we have shown above, the impurity potential couples to both spinons and rotors. However, the spinons are treated as free fermions with a bandwidth proportional to $Q_f$, which is quite small in the large-$U$ limit. Hence in the convolution, i.e. Eq. (\ref{eq:Gd-conv}), the spinon's spectral function acts approximately as a broadened $\delta$-function. The physical spectral function is dictated mostly by that of the rotor's, especially the Mott gap and consequently the in-gap spectra induced by impurities. Therefore, it is sufficient to focus on the impurity potential of the rotor fields. The impurity effects in the spinon sector can be safely ignored.
%Using the $T$-matrix approach, the
 {The}
full rotor Green's function, to the first order {in $h_0$}
% {IN WHAT?},
 % {ALSO, FOR ROTORS, IS THE IMPURITY WEAK OR STRONG?} \resp{the limiting value of $h_0 < 1$, and note this is single site comparing to the bulk state.}
  can be expressed as
  %{IS IT STILL $x0$ OR HAS IT BEEN SET TO THE ORIGIN?} \resp{I removed all specification of $\x_0$ as I actually carried $x_0$ throughout the manuscript.}
\begin{equation}
  \begin{split}
   % & \h{G}_X(\nu) %= ((\omega+ i \delta)^2 -\h{H}_0 -\h{H}_1 )^{-1}  \\
   % = \h{G}_{X,0}(\nu) + \h{G}_{X,0}(\nu) \h{T}(\nu) \h{G}_{X,0}(\nu),
%    & = (\omega -\h{H}_0 + i \delta)^{-1} + (\omega -\h{H}_0 + i \delta)^{-1} \h{T}
& G_X(\nu;\x_1,\x_2) = G_{X,0}(\nu;\x_1,\x_2) \\
& +  G_{X,0}(\nu;\x_1,\x_0) \meanvalue{\x_0}{\hat{T}}{\x_0} G_{X,0}(\nu;\x_0,\x_2).
  \end{split}
\end{equation}
Here, the rotor $T$-matrix is defined as
\begin{equation}\label{eq:T-matrix}
  \h{T} = \frac{\h{H}_{X,\x_0}}{1 - h_0 \nu G_{X,0}(\nu;\x_0,\x_0)},
\end{equation}
 where $\ket{\x}$ is the bulk rotor eigenfunction in the spatial representation.%, and $G_{X,0}(\nu;\x,\x') = \int \frac{d^2\k}{(2\pi)^2} e^{- i \k \cdot (\x - \x')}G_{X,0}(\nu;\k)$.
From here on, we abbreviate notations by writing $G_{X,0}(\nu;\x_i,\x_j) = g_{ij}$. Similarly, the retarded bulk rotor Green's function is written as $G^+_{X,0}(\nu;\x_i,\x_j) = g^+_{ij}$.
The impurity induced variation of the {  local rotor} spectral function { $\delta\rho_X(\nu;\x)  = \rho_X(\nu;\x) - \rho_{X,0}(\nu;\x)$} is derived from the retarded $\h{T}$ matrix:
\begin{align}\label{eq:delta-rho}
  & \delta\rho_X(\nu;\x_l) = -\frac{1}{\pi} \Im m [\frac{h_0 \nu g^+_{0l} g^+_{l0}}{ 1 - h_0 \nu g^+_{00} }].
\end{align}
It is convenient to separately discuss the two pieces of $\delta \rho_X(\nu;\x)$:
i) the first piece $\delta\rho_{X,1} (\nu;\x_0)$ comes from the original poles of $g^+_{00}$, i.e. the correction to the bulk Hubbard band, which reads
\begin{align}
   & \delta\rho_{X,1} (\nu;\x_0) =  \frac{2 h_0 \nu \rho_0(\nu) \Lambda_0(\nu) (1 - h_0 \nu \Lambda_0(\nu))}{(1 - h_0 \nu \Lambda_0(\nu))^2 + \pi^2 \rho_0(\nu)^2},
\end{align}
where $\Lambda_0(\nu) = \Re e[g^+_{00}]$ and $\rho_0(\nu) = -\Im m[g^+_{00}]/\pi$;
ii) the second contribution $\delta\rho_{X,2} (\nu;\x_0)$ comes from new poles that correspond to the vanishing denominator $1 - h_0 \nu g_{00}$. The new poles are only possible where $\rho_0(\nu;\x) \rightarrow 0$, i.e., for our concern, inside the Mott gap. $\delta\rho_{X,2} (\nu;\x_0)$ is expressed as
\begin{equation}\label{eq:d-rho-2}
  \begin{split}
  & \delta\rho_{X,2} (\nu;\x_0)  = \frac{h_0 \nu (\Lambda_0(\nu)^2  - \pi^2 \rho_0(\nu)^2)}{\abs{h_0( \nu \ \pt_\nu \Lambda_0(\nu) + \Lambda_0(\nu) )}} \delta(\nu-\nu_b).
  \end{split}
\end{equation}
The energy of the bound state $\nu_b$ is found by numerically {  solving the equation (see Fig. ({\color{blue}S1}) for numerical results of $g_{00}$)}
\begin{equation}\label{eq:vb}
1 - h_0 \nu g_{00} = 0.
\end{equation}
 Through Eq.~(\ref{eq:Gd-conv}), we see that $\nu_b$ determines the center of the electronic in-gap spectral weights.
%Therefore, we rewrite Eq. (\ref{eq:delta-rho}) as
%\begin{equation}
%  \label{eq:delta-rho-2}
%  \begin{split}
%   & \delta\rho_X(\nu;\x_l) = -\frac{h_0 \nu}{\pi}  \Big(\Im m[g^+_{l0} g^+_{0l}]  \Re e\left[(1 - h_0 \nu g^+_{00})^{-1}\right] \\
%   &+ \Re e[g^+_{l0}  g^+_{0l}] \Im m [(1 - h_0 \nu g^+_{00})^{-1}] \Big).
%  \end{split}
%\end{equation}

We show in Fig.~(\ref{fig:rotor-impurity}) the bulk rotor spectral function (red) and that on the impurity (blue). The Dirac-$\delta$ function is broadened as a Lorentzian.

The electronic spectral function obtained through {\dwx \eqdisp{eq:delta-rho}} is shown in Fig. (\ref{subfig:impurity-spectral}) and variation of the spectral function
% $ \Im m[\delta G^+_{d}(\omega; \x_0,\x_0)] =\Im m[G^+_{d}(\omega; \x_0,\x_0)]  - \Im m[G^+_{d,0}(\omega; \x_0,\x_0)] $
{  $\delta \rho_d(\omega;\x_0) = \rho_d(\omega;\x_0) - \rho_{d,0}(\omega;\x_0)$}
is given in Fig. (\ref{subfig:spectral-variance}). Both are in good agreement with
%that observed in
 {the experimental results of}
 Ref. \cite{Ye2012b}, which we
 %also
 quote in Supplemental Material \cite{SM}. Even though the experimental data are in arbitrary units, the relative area under the peak and above the dip in Fig. (\ref{subfig:spectral-variance}) is still quantitatively comparable to the experimental results.

%When measured away from the impurity site $\x \neq \x_0$, according to Eq.~(\ref{eq:delta-rho-2}), the energy of the bound state remains the same, but the hight, i.e. the spectral weight would be suppressed by the numerator of Eq.~(\ref{eq:d-rho-2}) when $\x$ is moving away from $\x_0$, in accordance with the experimental measurement shown in Fig.~(\ref{subfig:exp-spectral-variance}).

%{\color{red} * We also plot the variance of LDOS comparing the bulk}.
\begin{figure}
  \centering
   \subfigure[]{\label{subfig:impurity-spectral}\includegraphics[width=0.47\columnwidth]{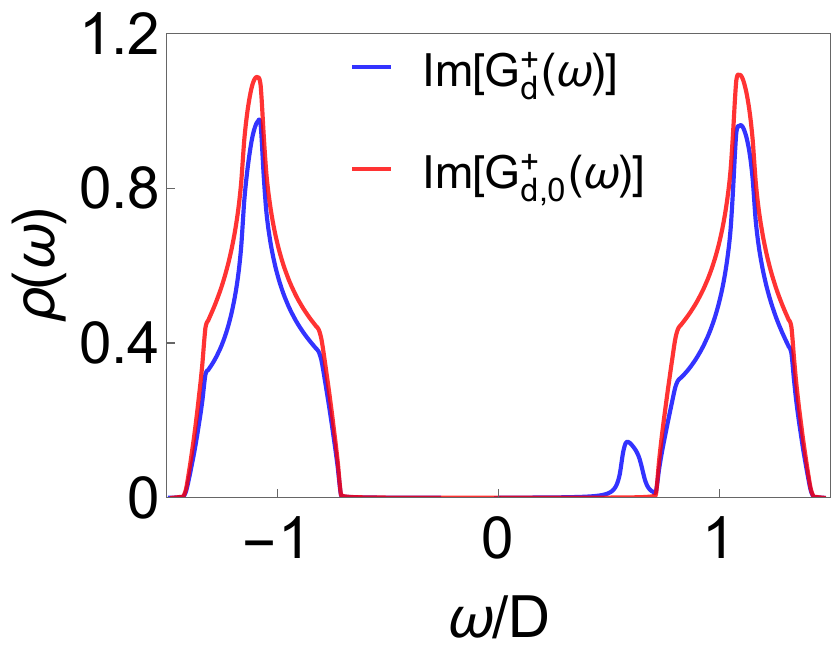}}
   \subfigure[]{\label{subfig:spectral-variance}\includegraphics[width=0.47\columnwidth]{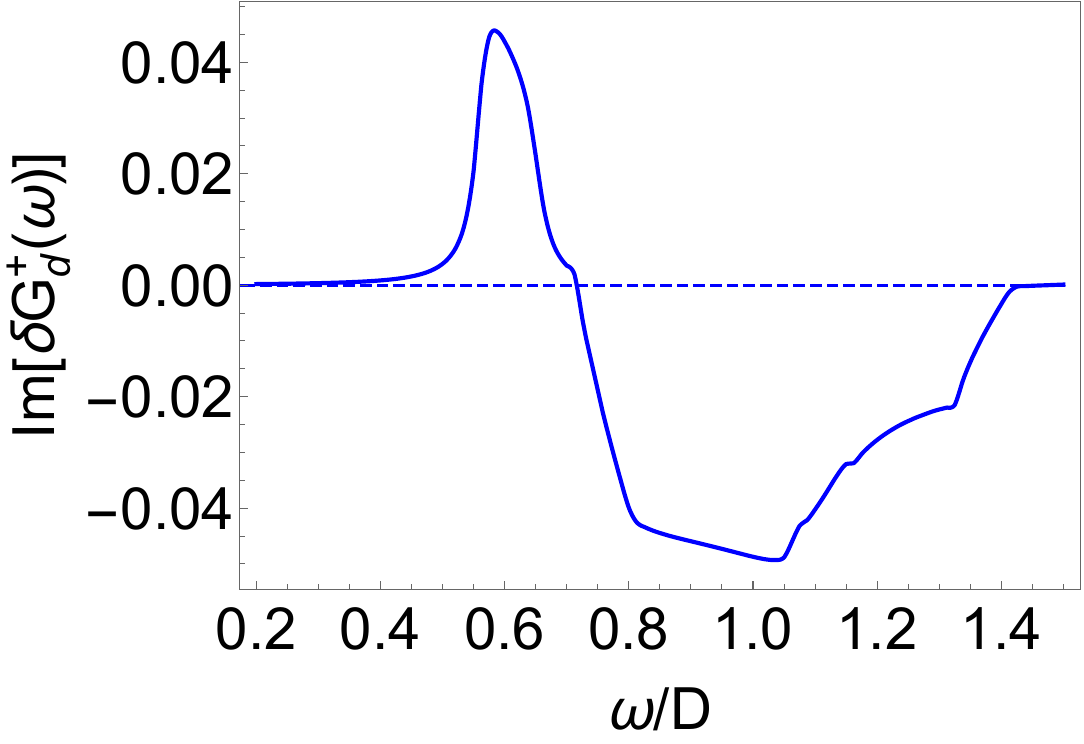}}
   \caption{(Color online) electronic LDOS on the impurity site (blue) compared with the bulk {  (red)} (\ref{subfig:impurity-spectral}), and the variation of  LDOS on the impurity(\ref{subfig:spectral-variance}).
   }\label{fig:electron-spectrum}
\end{figure}

{\it Solution for multiple impurities.}~~
The single impurity solution considered above is already the upper bound in terms of bound state energies, which are very close to the UHB. %For
 {However, in the experiments for} the Ca$^{2+}$
%cavity
 {vacancy}
of Ref.~\cite{Ye2012b}, the bound state is closer to the Fermi energy.
In the finite doping but still insulating cases \cite{Cai2015}, the binding energies of the in-gap states
% is
 {are}
 way beyond this limit.
 % and
  {They} can actually approach the Fermi energy but never reach it.
 %forming
  {The spectral weight of the in-gap states forms} a sharply $V$-shaped feature {centered} at the Fermi energy. We now show that, by considering multiple impurities, these properties are also captured within our framework: i) bound states of similar energies superimpose to create new bound states with smaller energies; ii) such bound states carry smaller spectral weights compared to that of the single impurity case.

For simplicity, we start with the two-impurity case.  We label
%the
 {their} positions
%of which we label
 as $\x_1$ and $\x_2$, and their corresponding single-impurity $T$-matrices as $\h{T}_1$ and $\h{T}_2$
 % as
  {which are similarly}
 defined
 in Eq. (\ref{eq:T-matrix}). To the lowest order, the full $T$-matrix is given by \cite{Economou2006}
\begin{equation}
  \label{eq:two-impurity-T-matrix}
  \h{T} = f_{12} (\h{T} _1 + \h{T}_2 + \h{T}_1 g_{12} \h{T}_2 + \h{T}_2 g_{21} \h{T}_1),
\end{equation}
with $f_{12} = (1 - t_1 t_2 g_{12} g_{21})^{-1},$ and $t_i = \nu h_0(\x_i) (1 - \nu h_{0}(\x_i) g_{ii})^{-1} $.
%First, both
 {Both} $\h{T}_1$ and $\h{T}_2$ contribute in-gap bound states at their own $\nu_{b,1(2)}$. Moreover, the factor $f_{12}$ contributes new bound states, the energies of which are deduced from
\begin{equation}\label{eq:2im-vb}
  \begin{split}
    & 1 - t_1 t_2 g_{12} g_{21} = 0 \Rightarrow \nu_b = \frac{A \pm \sqrt{A^2 - 4 B}}{B},
  \end{split}
\end{equation}
where $A = h_1 g_{11} + h_2 g_{22}$, $B = h_1 h_2 (g_{11} g_{22} - g_{12} g_{21})$, and $h_i = h_0(\x_i)$.
The new bound state solutions have the following properties: i) $\nu_b$ cannot reach zero just as for the single impurity case; ii) the new $\nu_b$'s are different from $\nu_{b,1}$ or $\nu_{b,2}$; iii) the new $\nu_b$'s are also positive-definite, meaning that they also descend down from the UHB; iv) most importantly, they become smaller, i.e. closer to the Fermi energy. In the single impurity case, the value of $h_0$ is bounded as $\abs{\delta \ev{n_f}} \le 1$ which further puts bounds on $\nu_b$ from below. However, in Eq. (\ref{eq:2im-vb}), when the two impurity potentials are close enough, we have $B \ll A$. By expanding Eq. (\ref{eq:2im-vb}) in terms of small $B$, we find
\begin{equation}\label{eq:new-nu_b}
\nu_b \sim 1/A = 1 / (h_1 g_{11} + h_2 g_{22}).
\end{equation}
Let $h_1 = h_2 = h_0$, and $g_{11} = g_{22} = g_{00}$, and compare Eq. (\ref{eq:new-nu_b}) with the single impurity case, $\nu_{b, \text{single}} \simeq g_{00} h_0$. Therefore, the new bound state can be considered as generated by an effective and larger $h'_0 \simeq 2 h_0$. In other words, the strength of impurities close together effectively adds up to produce bound states with lower and lower energies. Similar approximation can be made by considering the $T$-matrix of $M$ impurities (see Supplement Materials), i.e. when these impurities are sufficiently close, the bound state of  the lowest energy can be viewed as generated by a single impurity with all the impurity strength superimposed $h'_0 \simeq \sum_{m=1,\dots,M} h_m g_{mm}$, where the bound state energies $\nu_b \sim 1/h_0'$ are pushed closer to the Fermi energy as the number of impurities in a cluster increases.
%, comparing to the single impurity solutions which is considerably away from the Fermi energy.

%Therefore, we consider that the $h_0$ in the single impurity calculation can be tuned to larger values by nearby impurity concentration rather than bounded, and compute bound state energies and their corresponding spectral weights in the single impurity solution as a function of $h_0$.
{  Therefore, it is reasonable to use $h_0$, which now can go beyond the upper bound in the single impurity case, as a tuning parameter for impurity concentration. We shall compute the bound state energies and their corresponding spectral weights as a function of arbitrary $h_0$, and use the results to qualitatively explain experimental observation for doped samples.}
\begin{figure}
  \centering
  %\subfigure[]{\label{subfig:h0-omega-b}\includegraphics[width=4cm]{./omega_b-h0.pdf}}
  %\subfigure[]{\label{subfig:impurity-spectral-weight}\includegraphics[width=4cm]{./impurity-spectral-weight.pdf}}
  \subfigure[]{\label{subfig:imp-energy}\includegraphics[width=0.48\columnwidth]{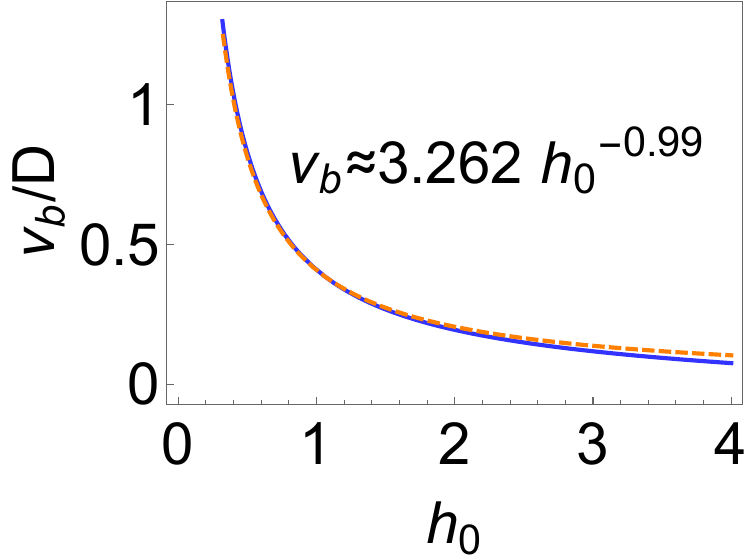}}
  \subfigure[]{\label{subfig:imp-sw}\includegraphics[width=0.48\columnwidth]{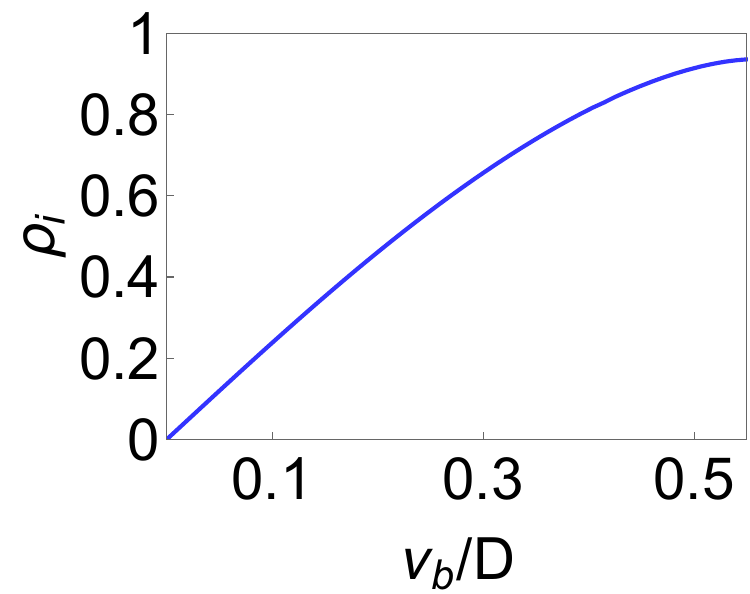}}
  \caption{The bound state energies $\nu_b$ plotted as a function of the effective rotor impurity potential $h_0$ (\ref{subfig:imp-energy}) and the spectral weights $\rho_i$ at given energy $\nu_b$ (\ref{subfig:imp-sw}).}\label{fig:omega-b-h0}
\end{figure}
We plot the energies of the in-gap states $\nu_b$ as a function of the impurity potential strength $h_0$ in {  Fig. (\ref{subfig:imp-energy})} and the corresponding spectral weights in Fig. (\ref{subfig:imp-sw}). Note that the asymptotic behavior is $\nu_b \sim 1/h_0$,  which forbids the impurity state
 {from}
reaching zero, i.e. the Fermi energy of the bulk. The corresponding spectral weight $\rho_i = \frac{(\Lambda_0(\nu_b)^2  - \pi^2 \rho_0(\nu_b)^2)}{\abs{ h_0 (\nu_b \ \pt_\nu \Lambda_0(\nu_b) + \Lambda_0(\nu_b))}} $ decreases when $\nu_b$ approaches zero.

{\it Experimental implications.}~~
From those results, we can explain the difference between the LDOS on the Cl$^{-}$ site and the Ca$^{2+}$ site.
The LDOS of Cl$^{-}$ vacancy is well explained by the single impurity solution, while that of a single Ca$^{2+}$ vacancy is much closer to the Fermi energy.
To explain the discrepancy, we note that the impurity potential of a Ca$^{2+}$ vacancy acts simultaneously on its {\em four} neighboring Cu sites (see Supplement Materials), forming a four-impurity cluster.
Thus, according to our approximation for the multi-impurity case, the peak position (relative to the Fermi energy) of Ca$^{2+}$ vacancy should be about $1/4$ of that for the Cl$^{-}$ site. The experimental results in Ref. [\onlinecite{Ye2012b}] give $\omega_b(\text{Cl}^{-}) \simeq 1.8 eV$ and $\omega_b(\text{Ca}^{2+}) \simeq 0.5 eV$, showing a quantitative agreement with our theory. A recent STM measurement \cite{wang2021} further confirmed our quantitative prediction with explicit comparison of measurements on single-impurity and two-impurity clusters.

{\it Discussion.}~~
Our work suggests that the observed V-shaped LDOS suppression is a generic feature of Mott insulators with or without magnetism. This is manifested by the special form of the effective rotor  impurity potential, $\h{H}_{X,\x_0} = i h_0(X^*_{\x_0}\pt_t X_{\x_0} - h.c.)$ which vanishes for $\nu = 0$ as long as the system is still in the Mott insulator phase.
More generally, the vanishing of LDOS reflects an exact zero of the local Green's function of the {\it parent Mott phase} rather than just the zeros of the spectral function, {as we show in the Supplement Materials.
This exact zero of the local Green's function is a consequence of the Luttinger surface of the electrons' Green's function in $\k$-space at the Fermi energy.}
%located  in $\k$-space
%% near
%at the spinon Fermi surface,
%where the
%% spinons have
% {spinon Green's function has}
%poles. These poles are turned into zeros through the convolution with the rotor's Green's function, which possesses a plane of zeros at $\nu=0$.
We propose that this Luttinger surface has topological stability related to the Mott gap, in a sense similar to the stability of a Fermi surface \cite{Volovik2009, Horava2005}. Exactly how such protection works is an intriguing open question for future studies.

The alignment of our theory with the experimental observations is also a demonstration of the emergence principle in the strongly correlated systems. Conventionally, the approach to such impurity problem would be implementing the $T$-matrix method directly with the physical Green's functions, such as in the simulation of Ref. \cite{Sun2019}. But such efforts
%all fail
have failed to explain the key aspects of the experimental observations.
{\dwx In our approach, the Luttinger point physics is captured through the proper treatment of the constraint to obtain the effective rotor impurity potential, \eqdisp{eq:imp-pot}.}
%Note that neither the local Luttinger point nor the $k$-space Luttinger surfaces were captured in the physical electronic Green's functions. Otherwise, a direct $T$-matrix computation with the electronic Green's functions would produce the same results. We attribute this is due to the approximation made in \eqdisp{eq:Gd}.}

While being able to account for many aspects of the impurity effects, our ultimate goal is to understand how finite concentration of impurities or holes eventually causes the Mott phase, %so as the
{\dwx and the concomitant} Luttinger surface, to collapse, which further leads to all these fascinating phenomena in the underdoped regime.
%However,
%While the slave rotor theory is known to be incapable of describing the finite doping phases,
%even at the mean field level,
%it
The slave rotor theory has the advantage of
%the explicit inclusion of
accounting for the Luttinger points or surfaces.
%is highly desirable
{\dwx Other
%similar
related
%slave-boson
methods,
%such as slave-spins,
even though may be capable of describing the finite doping phases, are not found to describe such impurity states. The current work suggests that future development of the formalism places a minimal requirement, viz.
%be consistent with the slave rotor theory in the half-filling limit as well as the impurity responses.
to capture the Luttinger surfaces of the half-filling limit.}
%In particular, the explicit inclusion of the Luttinger points or surfaces is highly desirable.
%The key, once again, is to be able to treat the Luttinger surfaces.}

{\it Summary.}~~
%In this work, we
We have studied the local density of states %LDOS
of a single impurity in the Mott insulator. %u
Using a slave rotor method,
 we solve the rotor impurity problem using the $T$-matrix method,
and find that the solution accounts for both of the key features of the observed
%LDOS
local density of states on a single impurity regarding i) positions and amplitudes of the in-gap spectral weights; ii) the spectral weight transfer from the UHB to the LHB.
Further analysis of the solutions for multiple impurities shows that high impurity concentration
%can induce
can be accounted for by a larger effective impurity potential
%$h_0$
for the rotor fields. We have emphasized that the Luttinger surfaces play an important role in our results.

We acknowledge useful discussions with Y. Y. Wang.
This work has primarily been supported by the NSF under Grant No. DMR-2220603,  and by the
Robert A. Welch Foundation Grant No. C-1411 and the Vannevar Bush Faculty Fellowship ONR-VB N00014-23-1-2870. W.D. was supported by the National Key R\&D Program of the MOST of China under Grant No. 2022YFA1602603 and the Startup Grant No. S020118002/002 of Anhui University.

% \bibliography{C:/Users/dwx0_000/OneDrive/PhyDir/Projects/library}
\clearpage
%%%%%%%%%%%%%%%%%%%%%%%%%%%%%%%%%%%%%%%%%%%%%%
\pagebreak
\newpage
% \onecolumngrid

\setcounter{figure}{0}
\makeatletter
\renewcommand{\thefigure}{S\@arabic\c@figure}
\setcounter{equation}{0} \makeatletter
\renewcommand \theequation{S-\@arabic\c@equation}
\renewcommand \thetable{S\@arabic\c@table}
\makeatletter
\def\maketitle{
\@author@finish
\title@column\titleblock@produce
\suppressfloats[t]}
\makeatother
\title{Supplemental Materials}
\maketitle

{\it The Slave Rotor Approach to Mott Insulators:~~}
Using $\pt_t \theta_i =  X_i^* \pt_t X_i$, we obtain the following Lagrangian from the original Hubbard model Hamiltonian:
\begin{equation}\label{eq:mf-L}
  \begin{split}
    & \L_{HM}  = \sum_{i,\alpha} f^\dagger_{i\alpha} (-i\pt_t  + h_i) f_{i\alpha} + \sum_i \Big(\frac{-\abs{\pt_t X_i}^2}{U}  \\
    & + \frac{h_i}{U}(X_i^* \pt_t X_i - X_i \pt_t X_i^*) + \lambda_i (\abs{X_i}^2 - 1)\Big)
\\ & +  \sum_{ij,\alpha} ( t_{ij} f^\dagger_{i,\alpha} f_{j,\alpha} X_i X^*_j + h.c.).
  \end{split}
\end{equation}
Note that $\frac{U}{2} \sum_{i} \hat{L}_{i}^2 =  \frac{-\abs{\pt_t X_i}^2}{2U}$; we have rescaled
 $U$ to $U/2$ in (\ref{eq:mf-L}) to preserve the correct atomic limit \cite{Florens2002}. Here, $h_i$ and $\lambda_i$ are two Lagrangian multipliers to enforce the two constraints
  {that were described in the main text}.

%The saddle-point solution \cite{Florens2004} corresponds to
%decoupling the
 {The} spinon-rotor coupling term
 {is decoupled} via
\begin{equation}
  Q_{f,ij}  = \ev{ X_j^* X_i}, \qquad
  Q_{X,ij}  = \ev{\sum_\alpha f_{j\alpha}^\dagger f_{i\alpha} } .
\end{equation}
The Lagrangian $\L_{HM}$ is {  expressed in} two parts:
\begin{equation}
  \label{eq:mf-Lf}
  \L_{MF,f} = \sum_{i,\alpha} (f^\dagger_{i\alpha} (-i\pt_t + h_i) f_{i\alpha} + t \sum_{\ev{ij},\alpha}
  (Q_{f} f^\dagger_{i\alpha} f_{j\alpha} + h.c.),
\end{equation}
\begin{equation}
  \label{eq:mf-LX}
  \begin{split}
    \L_{MF,X} & = \sum_i \Big(\frac{-\abs{\pt_t X_i}^2}{U} +
    \frac{h}{U}(X_i^*\pt_t X_i - X_i\pt_t X_i^* ) \\ &+ \lambda_i\abs{X_i}^2
    \Big) + t \sum_{\ev{ij}} (Q_{X} X_i X^*_j + h.c.).
  \end{split}
\end{equation}
%This gives rise to a
A saddle-point solution
 {arises}
when one generalizes each $X_i$ to $M$ species so that its symmetry becomes $O(2M)$,
scales the hoping
$t_{ij}$ to $1/M$, and takes the large (N,M) limit with a fixed ratio $M/N$ \cite{Florens2004, Lee2005}. In our analysis,
we
%will write
{express} our equations for $N=2M=2$.

{\dwx Numerically, we choose $U=16 t$ which is about $1.5~U_c$ in two-dimensions within the slave rotor theory. $t Q_f \simeq 0.3$ at this point while $t Q_X \simeq 0.8 t$ disregarding the value of $U$. Then solving \eqdisp{eq:mf-Lf} and (\ref{eq:mf-LX}) yields
  \begin{align}
  G_{f,0}(\omega;\k) = (\omega + h -Q_f \epsilon_{\k})^{-1}, \\
 G_{X,0}(\nu;\k) = (- \nu^2/U + 2ih\nu/U + \lambda + Q_X \epsilon_{\k})^{-1},
\end{align}
}

{\dwx
  In Fig. (\ref{sfig:GX-tordered}), we show the real part ($\Re e [G_{X,0}(\nu; x_0, x_0)]$, blue) and the imaginary part ($\Im m [G_{X,0}(\nu; x_0, x_0)]$, red) of the time-ordered, saddle point rotor Green's function $G_{X,0}(\nu; x_0, x_0)$ which are used in the solving for the rotor $\hat{T}$-matrix.
  \begin{figure}
    \centering
    % \subfigure[]{\label{fig:rotor-bulk}\includegraphics[width=0.85\columnwidth]{./G0-spectrum.pdf}}
    \includegraphics[width=0.9\columnwidth]{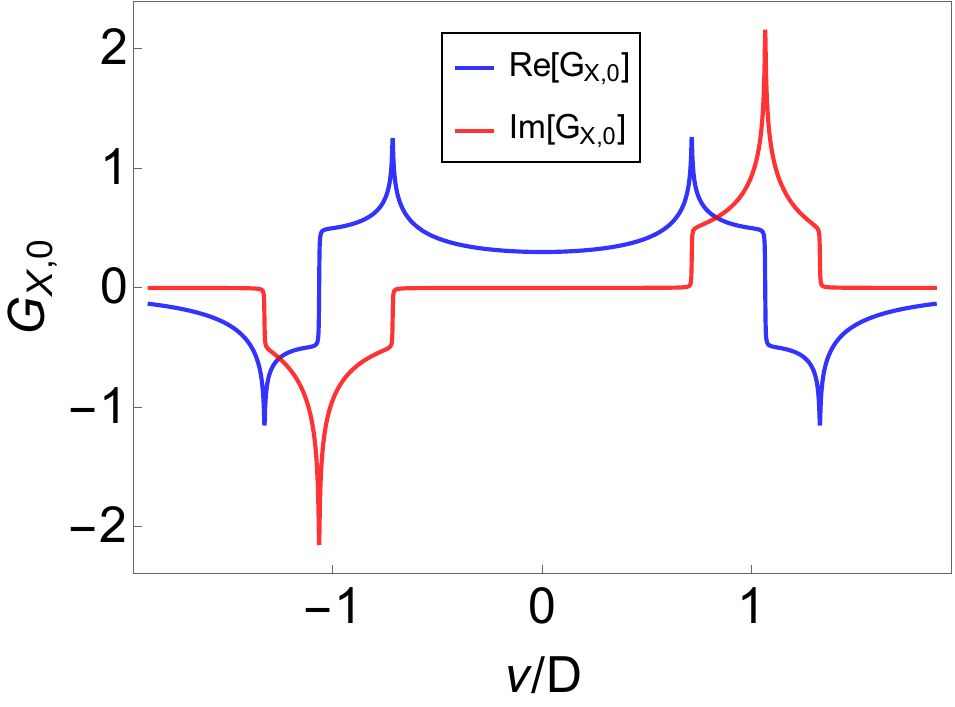}
    \caption{(Color online) (\ref{sfig:GX-tordered}) the real part ($\Re e [G_{X,0}(\nu; x_0, x_0)]$, blue) and the imaginary part ($\Im m [G_{X,0}(\nu; x_0, x_0)]$, red) of the time-ordered saddle point rotor Green's function $G_{X,0}(\nu; x_0, x_0)$.}
    \label{sfig:GX-tordered}
\end{figure}
}

%%%%%%%%%%
{\it Calculation of the constraint:}~~
The electron density is given by $\ev{\sum_{\sigma}d^\dagger_\sigma d_\sigma} =  \ev{\sum_{\sigma}f^\dagger_\sigma f_\sigma}$, and the rotor fields are constrained by $\ev{\hat{L}} = \ev{U^{-1} (  - i \pt_t \theta + h) } = \ev{\sum_\sigma f^\dagger_\sigma f_\sigma -1/2}$. To enforce the constraints, we need to first compute $\ev{\hat{L}}$ at finite $h$. Since $X_i = e^{i \theta_i}$, we have $\theta_i = -i \ln X_i$, therefore, $   - i \pt_t \theta_i = - X_i^* \pt_{t} X_i$,
 {and}
%Note that we can also choose $   - i \pt_t \theta_i = X_i \pt_{t} X^*_i$, which is a consequence of the constraint between $X$-field and $\theta$-field. It is a natural choice to select an explicit Hermitian form in terms of the $X$-field:
\begin{equation}
  \hat{L} = (-  X_i^* \pt_{t} X_i + h)/U.
\end{equation}
%Then we use
 {We compute $\ev{\hat{L}}$ using}
the functional integral
% to compute $\ev{\hat{L}}$
 in the $X$-field representation
  {and find}
 %:
\begin{equation}
  \begin{split}
    & \ev{\hat{L}} %=  h/U  + \int \frac{d\nu}{2 \pi} \frac{ i \nu / U}{-\nu^2/U + \lambda - 2 i h \nu/U} \\
    =  h/U - \frac{h/2}{\sqrt{- h^2 + \lambda U}}.
  \end{split}
\end{equation}
%Note that we do not take derivative on the $\lambda X^2$ term. The $\delta \tau$ is introduced for regularization purpose.
When there is no doping, it follows from Eq. (2) that $\ev{\hat{L}} = 0$.

\begin{figure}[h]
  \centering
     \subfigure[]{\label{subfig:exp-impurity-spectral}\includegraphics[width=0.45\columnwidth]{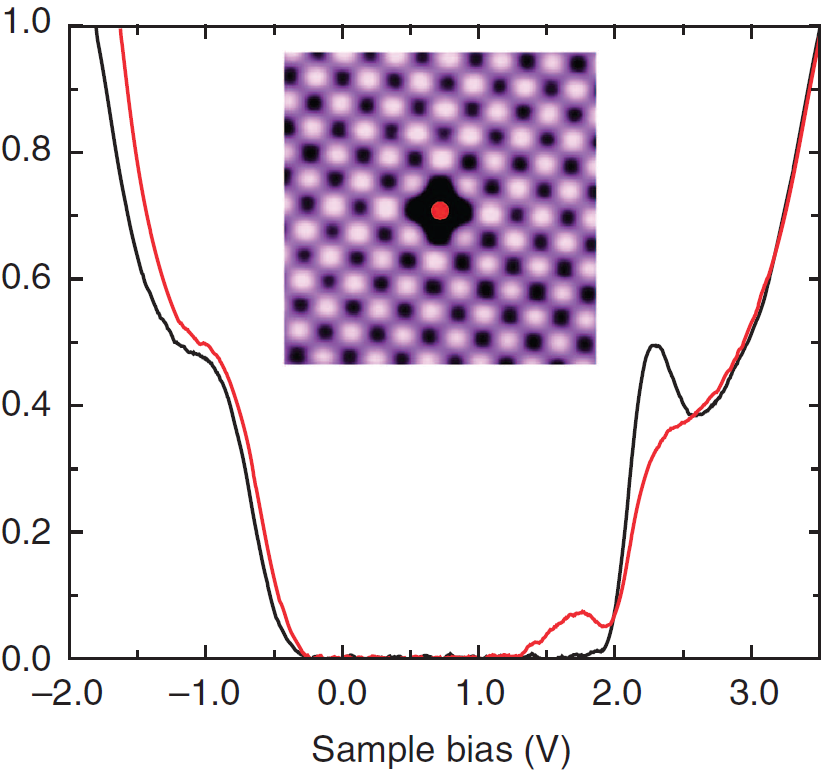}}
     \subfigure[]{\label{subfig:exp-spectral-variance}\includegraphics[width=0.45\columnwidth]{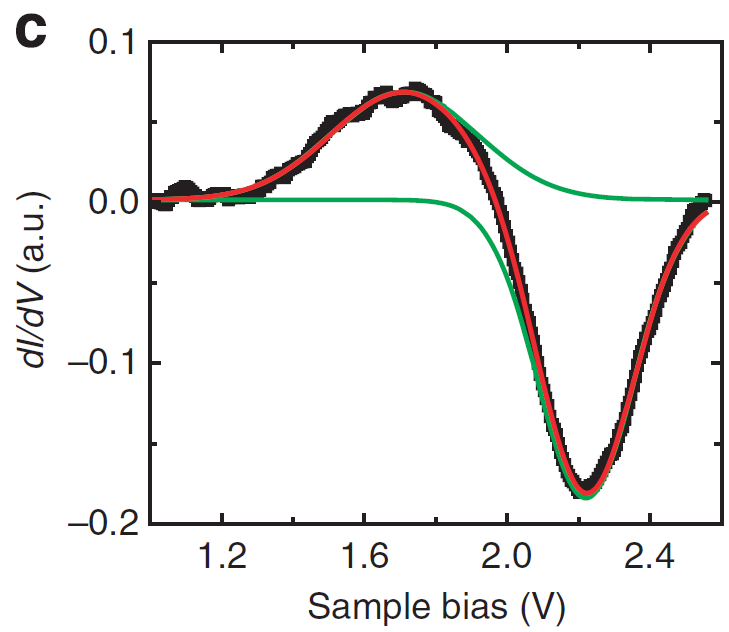}}
     \subfigure[]{\label{subfig:exp-cluster-spectral}\includegraphics[width=0.45\columnwidth]{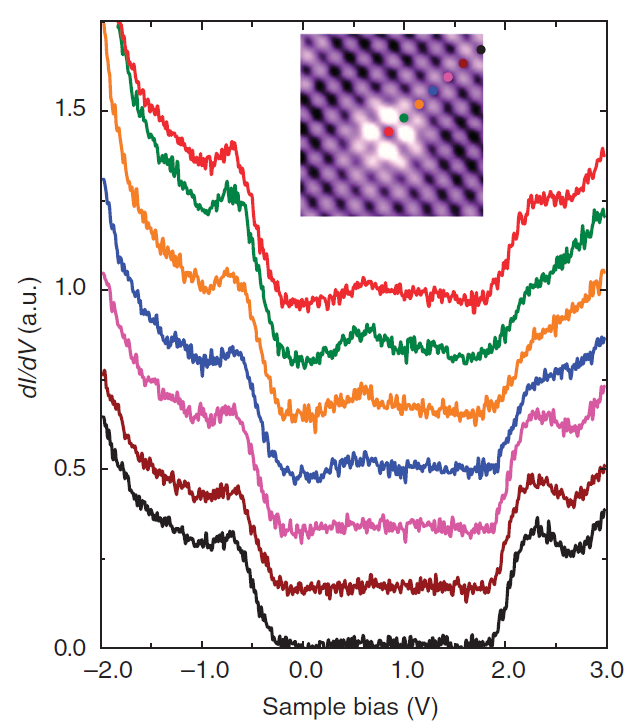}}
     \subfigure[]{\label{subfig:imp-pos-cl}\includegraphics[width=0.45\columnwidth]{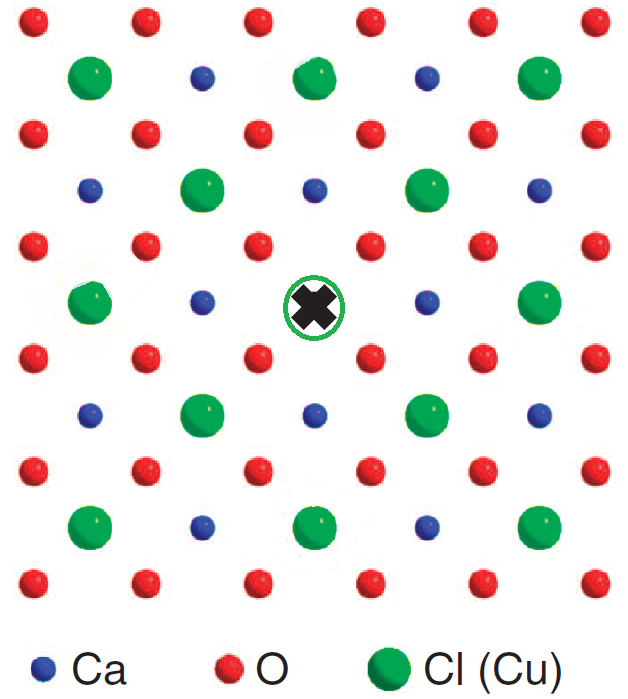}}
     \subfigure[]{\label{subfig:imp-pos-ca}\includegraphics[width=0.45\columnwidth]{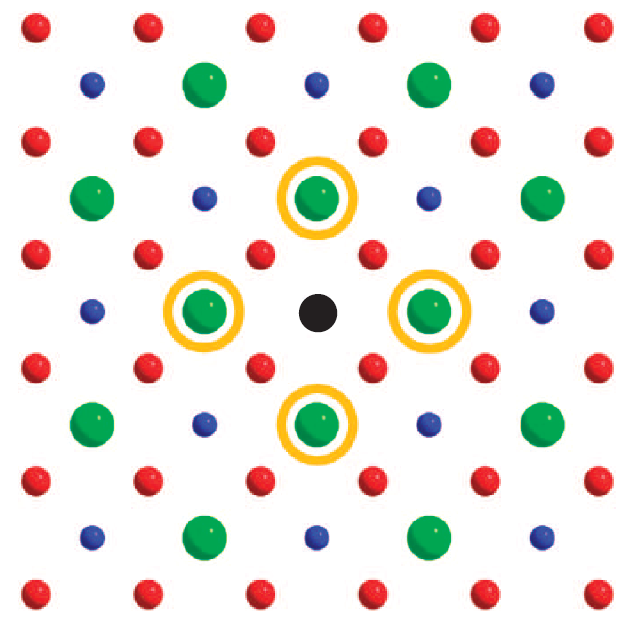}}
     \subfigure[]{\label{subfig:exp-m-imp}\includegraphics[width=0.52\columnwidth]{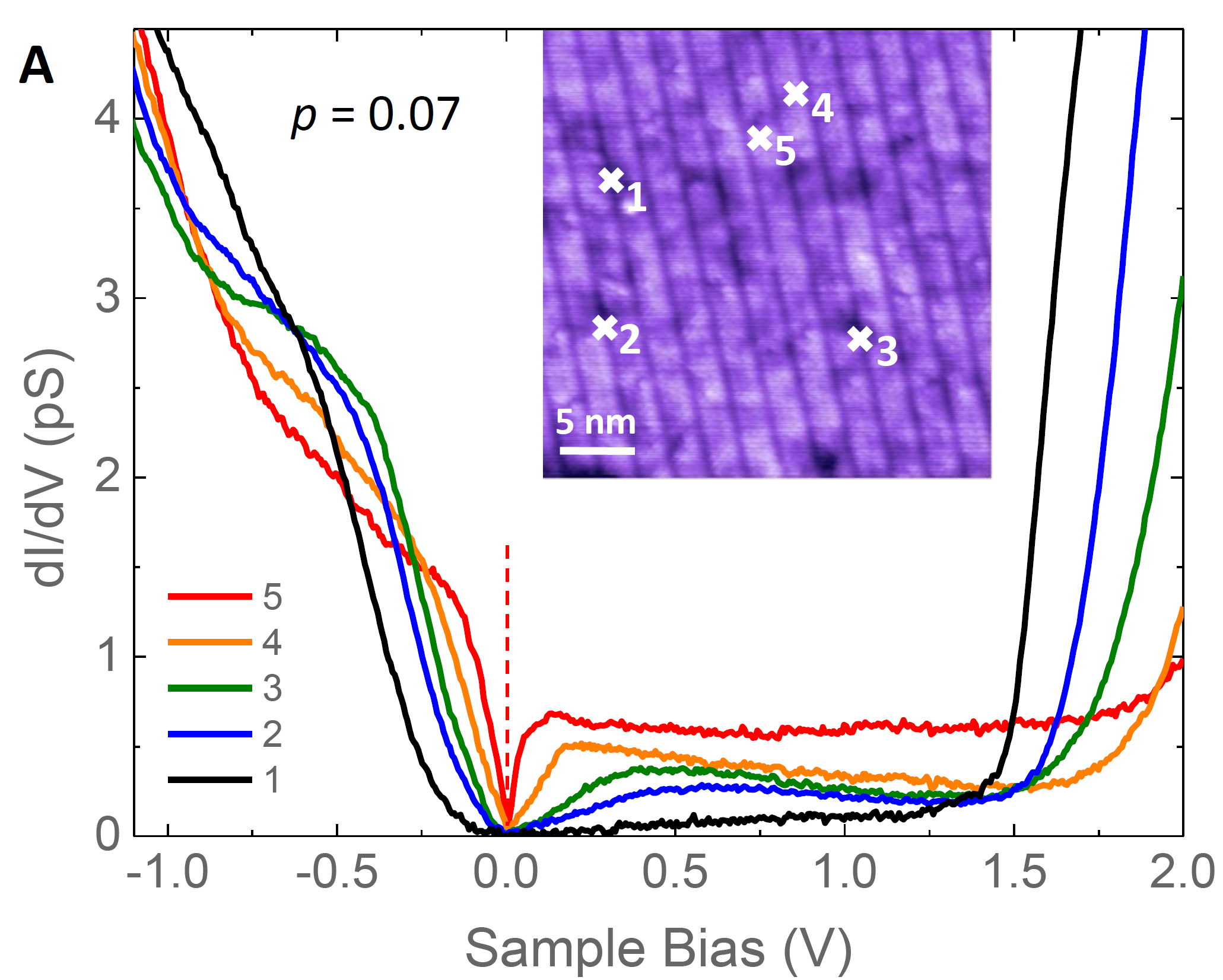}}
     \caption{(a) Single impurity LDOS (red) compared with the bulk LDOS (black).
      {The single impurity corresponds to a Cl$^{-}$ vacancy.}
     (b) Difference of LDOS between the impurity site and the bulk state. (c) Measurement on a Ca$^{2+}$ vacancy, where the potential effectively forms a four-impurity cluster.
     % and forms
      {The} bound state  {occurs}
     at energy much closer to the Fermi energy compared to that of the Cl$^{-}$ vacancy in (a). (d) Schematic topview of exposed surface showing the missing Cl defect (dark cross) which is on top of a Cu atom (green circle). (e) Schematic top view showing the position of the Ca-site defect (dark dot) and the nearest-neighbour Cu sites (orange circles). (f) The LDOS of a 7\% doping sample which shows a sharp ``V''-shaped dip.}
    \label{sfig:1}
  \end{figure}

{\it Experimental measurements.}
Experimental results of
 {the} impurity LDOS in cuprate parent compounds
 {are reproduced}
% shown
in Fig.~(\ref{sfig:1}).

  {\it Multi-impurity T-matrix:}~~
 Consider $M$ impurities labeled by $\x_m$ with $m = 1, 2, \dots, M$. The T-matrix is now given by
\begin{equation}
  \h{T} = (I_M - \h{V} \h{\Gamma})^{-1} \h{V},
\end{equation}
where $I_M$ is an identity matrix of dimension $M\times M$, $\h{V}$ is a diagonal matrix with elements $V_{mm} = h_0(\x_m)$, and $\h{\Gamma}_{ij} = g_{ij}$ with $\x_{i(j)}$ belonging to the impurity sites. The new poles are solved from $\det[I_M - \h{V} \h{\Gamma}] = 0$.
By keeping to second order terms ($\sim \mathcal{O}(\nu^2)$), we find $\nu_b \sim (\sum_m h_m g_{mm})^{-1}$ with $\x_m$ denoting positions of the impurities.

  {\it Probing the zero of local Green's function:}~~
{To further shed light on interpretation of the experimental results, we propose that}
we can interpret the impurity LDOS as a measurement of the local and energy resolved compressibility of the bulk state as follows. {The local impurity potential distribution of $\mu_{imp}(\x)$ within a sample can be viewed as an ensemble of source fields $\{\delta \mu(\x_0)\}$ that couples to the density operator. Then $\rho^{imp}(\omega;\x)$ is equivalent to $\rho(\omega;\x;\{\delta \mu(\x_0)\})$, the LDOS in the presence of a source field ensemble $\{\delta \mu(\x_0)\}$. Theorerically, we consider the source field ensemble $\{\delta \mu(\x_0)\}$ tunable. Experimentally, the system is well within the Mott insulating phase, and no phase separation appears. Therefore, the variation of LDOS due to the source fields:
  \begin{equation}
  \delta \rho(\omega; \x; \{\delta \mu(\x_0)\})= \rho(\omega;\x;\{\delta \mu(\x_0)\}) - \rho^{bulk}(\omega;\x)
\end{equation}
  can be further considered as linear response, i.e. we can take the zero source limit $\{\delta \mu(\x_0)\} \rightarrow 0$. Such linear response can be formally expressed as:}
%   {CAN YOU ELABORATE ON THIS EQUATION?}:
 { \begin{equation}\label{sup-eq:rho-imp-var}
    \begin{split}
      & \left. \frac{\delta \rho(\omega; \x; \{\delta \mu(\x_0)\})}{\{\delta \mu(\x_0)\}}\right\vert_{\{\delta\mu(\x_0)\}\rightarrow 0}  = \frac{\delta \rho^{bulk}(\omega; \x)}{\delta \mu},
  \end{split}
\end{equation}
and the measured LDOS variation is expressed as
\begin{equation}
\delta \rho(\omega; \x; \{\delta \mu(\x_0)\}) \simeq \frac{\delta \rho^{bulk}(\omega; \x)}{\delta \mu} \times  \{\delta \mu(\x_0)\}.
\end{equation}
Therefore, the key experimental observation that no spectral weight of LDOS in the presence of impurities is allowed at $\omega = 0$ can be viewed as
\begin{equation}
  \frac{\delta \rho^{bulk}(\omega = 0; \x)}{\delta \mu} = 0,
\end{equation}
as $ \{\delta \mu(\x_0)\}$ is arbitrary (within the linear response region).}

Then we can further transform $\frac{\delta \rho^{bulk}(\omega; \x)}{\delta \mu}$ as
    \begin{equation}
      \frac{\delta \rho^{bulk}(\omega; \x)}{\delta \mu} = \frac{\pt \rho^{bulk}(\omega; \x)}{\pt \mu} (-\frac{\delta \Sigma(\omega;x)}{\delta \mu} - 1),
\end{equation}
where $\Sigma(\omega;x)$ is the local electronic Dyson self-energy. Assuming that $$(-\frac{\delta \Sigma(\omega = 0;x)}{\delta \mu} - 1) \neq 0 $$ which is generally true, the vanishing of the LDOS means that
\begin{equation}
  \frac{\pt \rho^{bulk}(\omega = 0; \x)}{\pt \mu} = 0.
\end{equation}
 Then we can write the real part of the Green's function through the Hilbert transform as
  \begin{equation}
    \Re e[G(\omega; \x)] =  \mathcal{P} \int d\omega' \frac{\rho^{bulk}(\omega'; \x)}{\omega - \omega'}.
  \end{equation}
  Taking $\omega = 0$, the integral on the right hand side is completely determined by the $\omega' = 0$ contribution as the contribution from finite $\omega'$ cancels due to dynamic particle-hole symmetry of the bulk state spectral function {$\rho^{bulk}(- \omega'; \x) = \rho^{bulk}(\omega'; \x)$ (which is only true at half filling)}
  \begin{equation}
    \begin{split}
      & \Re e[G(\omega = 0; \x)] =  -\mathcal{P} \int d\omega' \frac{\rho^{bulk}(\omega'; \x)}{\omega'} \\
      & = - \frac{\pt \rho^{bulk}(\omega'; \x)}{\pt \omega'} \vert_{\omega' \rightarrow 0}\\
      & \propto \frac{\pt \rho^{bulk}(\omega = 0; \x)}{\pt \mu} = 0,
   \end{split}
 \end{equation}
 where in the last line we used the relation $$\frac{\pt \rho^{bulk}(\omega; \x)}{\pt \mu} \propto \frac{\pt \rho^{bulk}(\omega; \x)}{\pt \omega}$$
 as $\mu$ only explicitly shows up in electronic Green's function as
 \begin{equation}
   G(\omega; \x) = \int d^2\k \frac{e^{i \k \cdot \x}}{\omega - \mu - \epsilon_{\k} - \Sigma(\omega; \k)}.
 \end{equation}

% \bibliography{C:/Users/dwx0_000/OneDrive/PhyDir/Projects/library}
 %

 %----------------------------------------------------------%
\end{document}